\documentclass{aa}  
\usepackage{graphicx}
\usepackage{txfonts}
\usepackage[]{hyperref}
\usepackage{xcolor}
\usepackage[normalem]{ulem}
\usepackage{subcaption}

\urldef{\stixflarelist}\url{https://github.com/hayesla/stix_flarelist_science}
\urldef{\stixdatacenter}\url{https://datacenter.stix.i4ds.net/}
\urldef{\ospex}\url{https://hesperia.gsfc.nasa.gov/ssw/packages/spex/doc/ospex_explanation.htm}
\urldef{\solarorbiterarchive}\url{https://soar.esac.esa.int/soar/}
\urldef{\sunkitspexstix}\url{https://github.com/natsat0919/sunxspex/tree/albedo-correction}

\begin{document} 

    \title{Spectral component imaging of solar X-ray flares}

   \author{Muriel Zo\"{e} Stiefel
          \inst{1}\fnmsep\inst{2}
          \and
          Paolo Massa \inst{1}
          \and
          Alessia Guidetti \inst{3}
          \and
          Marina Battaglia \inst{1}
          \and
          S\"{a}m Krucker\inst{1}\fnmsep\inst{4}
          }

   \institute{University of Applied Sciences and Arts Northwestern Switzerland, Bahnhofstrasse 6, 5210 Windisch, Switzerland\\
              \email{muriel.stiefel@fhnw.ch}
        \and
            ETH Z\"{u}rich, 
            R\"{a}mistrasse 101, 8092 Z\"{u}rich Switzerland
        \and
            MIDA, Dipartimento di Matematica, Università di Genova, via Dodecaneso 35, 16146 Genova, Italy
        \and
            Space Sciences Laboratory, University of California, 7 Gauss Way, 94720 Berkeley, USA
            }

   \date{Received September 23, 2025; accepted November 14, 2025}

  \abstract
   {Solar hard X-ray (HXR) observations provide diagnostics of the hottest plasmas and of nonthermal electron populations present during solar flares and coronal mass ejections. HXR images of specific energy ranges often contain overlapping contributions of these components, complicating their interpretation. This is even more challenging as HXR imagers generally use an indirect imaging system.}
   {Our work aims to separately image individual spectral components, such as thermal loops, superhot sources, and nonthermal footpoint sources, rather than obtaining images of specific energy ranges that show a combination of all components.}
   {We introduce a new method called ``spectral component imaging'' and apply it to observations provided by the Spectrometer/Telescope for Imaging X-rays (STIX) aboard Solar Orbiter. First, the flare integrated HXR spectrum is fitted with individual spectral components to get the relative contributions (``weights'') of each component in each native STIX energy channel. In a second step, a set of linear equations is created based on these weights and the observed, energy-dependent STIX visibilities. The visibilities of the individual spectral components are derived by means of a linear least-squares approach and are subsequently utilized for image reconstructions.}
   {We demonstrate the effectiveness of spectral component imaging on four different flares observed by STIX. This method provides powerful diagnostics, particularly for flares with hot and superhot components, allowing us to spatially separate these two thermal components. We apply our methodology to the nonthermal peak of the X7.1 flare SOL2024-10-01, and we find that the superhot component is located 4.8 Mm away from the hot thermal loops. The thermal energy of the superhot component is approximately 20\% of the energy content of the hot component, highlighting the significance of superhot components in the total flare energy budget.}
   {Spectral component imaging provides a powerful tool to image individual spectral components (i.e., thermal and nonthermal X-ray sources), rather than creating images over fixed energy ranges. Because there is no need to select an energy range, spectral component imaging has the potential to automate the image reconstruction process and to establish a robust STIX image database once the spectral components have been defined.}


   \maketitle

\section{Introduction}\label{Intro}

    A solar flare occurs when there is a rapid release of magnetic energy in the solar corona, caused by the rearrangement of magnetic field lines into a more relaxed configuration. Flares are visible over the entire electromagnetic spectrum as a sudden brightening that can last from minutes to several hours \citep[e.g.,][]{Fletcher_2011, Benz_2017}. Emissions observed in the hard X-ray (HXR) range are produced by the bremsstrahlung mechanism, a mechanism well understood and quantified \citep[e.g.,][]{Brown_1971}. This makes HXR measurements one of the most direct probes of the energy release process in solar flares. Only the most energetic electrons in the solar atmosphere are capable of producing bremsstrahlung in the HXR range ($\gtrsim$ 4 keV). Thus, HXR instruments are ideal for measuring the hottest thermal plasma ($\gtrsim$ 8 MK) and the distribution of nonthermal energetic electrons. In the context of the standard flare model \citep[e.g.,][]{Shibata_1995}, nonthermal electrons, accelerated in the solar corona due to energy release in the process of magnetic reconnection, precipitate along magnetic field lines into the denser chromosphere, where they interact with the surrounding plasma and produce bremsstrahlung. The bremsstrahlung emission from nonthermal electrons is best described using a power-law distribution based on the thick target model \citep[e.g.,][]{Brown_1971}. The nonthermal emission typically exhibits a morphology characterized by footpoint sources, which correspond to the locations where the majority of nonthermal electrons collide with the chromosphere. This interaction heats the chromosphere, resulting in chromospheric evaporation \citep[e.g.,][]{Antonucci_1982}. Chromospheric material evaporates into the corona, leading to the formation of hot flare loops that can be observed in HXR, soft X-ray (SXR), and the extreme ultraviolet (EUV) range. The spectrum of these hot flare loops observed by HXR instruments is typically approximated by a single thermal Maxwellian electron distribution, even though flares are multi-thermal in nature \citep[e.g., ][]{Warren_2013}.

    Spectral analysis of large flares ($\gtrsim$ X1) in the HXR range revealed that a single isothermal model is often not appropriate to obtain a good fit to the data \citep[e.g., ][]{Caspi_2010}. Therefore, adding a second thermal component is necessary to improve the spectral fit. The two components were named the hot thermal component, typically with temperatures below 30 MK, and the superhot thermal component, characterized by temperatures exceeding 30 MK \citep[e.g.,][]{Lin_1981, Hudson_1985, Tanaka_1987}. The superhot component is regularly observed in HXR spectral analyses \citep[e.g.,][]{Caspi_2014, Nagasawa_2022, Stiefel_2025b}. The origin of the superhot component is currently still debated. It could be the hottest thermal loop produced by evaporated chromospheric plasma, or, more commonly suggested, it may be a signal of direct heating of coronal plasma associated with the reconnection process \citep[e.g.,][]{Longcope_2011}. The lack of understanding about the superhot component's origin stems largely from the challenges associated with imaging it. In \citet{Caspi_2015}, Reuven Ramaty High-Energy Solar Spectroscopic Imager \citep[RHESSI; ][]{Lin_2002} visibilities from two specific energy ranges were used to retrieve the visibilities of the two thermal components for a single flare. The location of the superhot source was found to be clearly above the hot flare loops, favoring the direct heating explanation. Due to the high temperature and low emission measure of the superhot component \citep[e.g.,][]{Stiefel_2025b}, EUV and SXR instruments are hardly sensitive to it. Their measurements are dominated by the emission from cooler flare loops. Therefore, HXR instruments are the ideal candidates for imaging the superhot component. 

    The currently operating solar-dedicated HXR imagers are the Spectrometer/Telescope for Imaging X-rays \citep[STIX; ][]{Krucker_2020} on Solar Orbiter \citep{Muller_2020} and the Hard X-ray Imager \citep[HXI; ][]{Zhang_2019} onboard the Advanced Space-based Solar Observatory \citep[ASO-S; ][]{Gan_2023}. Both instruments utilize an indirect imaging system based on a Fourier-transform imaging technique. When reconstructing images, users must make several decisions regarding, e.g., the time range, the energy range, and the detectors to be considered. The standard method for HXR image reconstruction involves generating images of the thermal and the nonthermal emissions by utilizing specific energy ranges in which one component dominates, ignoring the energy ranges with overlapping spectral contributions. Typically, thermal emissions dominate at lower energies, while nonthermal emissions dominate at higher energies. The transition between these two components varies depending on the flare size and can be identified through spectral analysis. Typically, for smaller flares this transition occurs at lower energies \citep[e.g.][]{Hannah_2008}. This standard imaging method is effective when the flare has clear energy ranges with high enough counting statistics that are dominated by individual spectral components. 

    The standard imaging method becomes more challenging in the extreme case of solar flares observable with HXR instruments, including microflares and large X-class flares. Imaging the nonthermal component of microflares is challenging because there is no energy range with sufficient counting statistics where the nonthermal component clearly dominates without the influence of thermal emission. For large flares, two main issues arise. First, HXR instruments typically use an attenuator during periods of high flux levels to block low-energy emissions, which reduces the number of energy ranges available for reconstructing thermal images. Second, thermal emissions are often fitted using two thermal components, making it difficult to find energy ranges where the emission from one thermal component distinctly outweighs that of the other. 
    
    In this paper, we introduce a method for imaging individual spectral components derived from previously established spectral fits, which we refer to as ``spectral component imaging''. It is important to differentiate this method from ``imaging spectroscopy''. In the latter, the flux from individual sources (e.g., two separate footpoint sources) as a function of energy is extracted from energy-dependent images. This allows for a comparison between the spectra of the individual sources \citep[e.g.,][]{Krucker_2002}. A detailed discussion on the conceptual differences between imaging spectroscopy and spectral component imaging is in Section \ref{IS vs SCI}.
    
    The fundamentals of spectral component imaging are based on \citet{Caspi_2015}. Therefore, we will first discuss the key points of \citet{Caspi_2015} before introducing our new approach in Section \ref{Section: Method}. Unlike the \citet{Caspi_2015} method, we suggest utilizing all available energy channels from the HXR instrument, building an overdetermined linear system that can be solved analytically with the least-squares approach. In Section \ref{Section: Method}, we will outline the differences between the two methods in more detail. In Section \ref{Applications}, we will demonstrate that spectral component imaging is effective with STIX data and we will present various applications of the method. Finally, in Section \ref{Discussion}, we will discuss the advantages and limitations of spectral component imaging, highlight the distinctions from imaging spectroscopy, and draw our conclusions.

\section{Methodology: Spectral component imaging} \label{Section: Method}

    In the following Section, we discuss the mathematical principles of spectral component imaging. The general mathematical concept is applicable to any instrument that uses visibilities. In the second part, we outline how spectral component imaging is utilized with STIX data.

    \subsection{Mathematical concept of spectral component imaging using visibilities}\label{Math}

    Rather than having visibilities for a specific energy range, spectral component imaging aims to determine the visibilities of the different spectral components (e.g., thermal and nonthermal components). In the following, we will discuss the mathematical concept of spectral component imaging. The first part of the description follows \citet{Caspi_2015}.

    The primary simplifying assumption is that the morphology of the spectral component to be imaged (called source A in the following) is energy independent. However, the contribution of source A to the total measured spectrum of a flare varies with energy. Consequently, when two sources, A and B of different origins, contribute to the total spectrum, images created in different energy bands will appear distinct due to varying contributions from sources A and B. This concept is illustrated in Fig. 1 of \citet{Caspi_2015}.

    The assumption that the source morphology is energy-independent is comparable to the simplification using an isothermal fit or two thermal fits in HXR spectral fitting. While these simplifications can not account for the full complexity of a solar flare, they are reasonable given the limitations and dynamic range of the HXR instruments. If the source morphology or the contribution of the different spectral components changes during the considered time interval, the method can only capture their time-averaged morphology and spectrum. Therefore, these assumptions effectively describe the average behavior of flares.

    We want to point out two important points regarding visibilities \citep[following][]{Caspi_2015}.
    First, a visibility $V$ observed at energy $E$ and corresponding to an angular frequency $(u,v)$ can be expressed as the product of the total integrated flux $F$ and the relative visibility $\nu$ containing information on the morphology of the source(s), i.e.
    \begin{equation}
        V(u,v,E) = F(E)~\nu(u,v,E) ~.
        \label{Eq: Visibility flux/morphology }
    \end{equation}
    Second, if two (or more) spectral sources contribute to a visibility measured in an energy range $E$, the visibility $V(E)$ can be written as the sum of all the visibilities of the individual sources, i.e.
    \begin{equation}
    V(u,v,E) = \sum_i V_i(u,v,E)~.
    \label{Eq: Visibility linearity}
    \end{equation}
    By applying Eq. \eqref{Eq: Visibility flux/morphology } to both the left- and right-hand side of Eq.~\eqref{Eq: Visibility linearity}
    we obtain
    \begin{equation}
        F(E)~\nu(u,v,E) = \sum_i F_{i}(E)~\nu_{i}(u,v,E) \approx \sum_i F_{i}(E) ~\nu_{i}(u,v)~,
        \label{Eq: Visibilities as sum}
    \end{equation}
    where we use the assumption that the morphology of the individual components $\nu_{i}(u,v)$ are not energy-dependent.
    Dividing Eq.~\eqref{Eq: Visibilities as sum} by $F(E)$, we get
    \begin{equation}
        \nu(u,v,E) = \sum_i f_{i}(E) ~\nu_{i}(u,v),
        \label{Eq: Visibilities as sum fraction}
    \end{equation}
    where $f_i(E) = F_i(E)/F(E)$ are the fractional contributions of each source to the total measured flux. Spectral analysis provides the fractions $f_i(E)$ for each source as a function of energy. Spectral component imaging aims to solve for the visibilities $\nu_{i}(u,v)$, which contain the morphology of the spectral components.
    
    The approach of \citet{Caspi_2015} is to have $N$ equations (corresponding to energy ranges $E_1, \dots, E_N$) for $N$ unknown spectral X-ray sources. Therefore, when using two components (for instance, thermal and nonthermal) for spectral analysis, the user must select two energy ranges, build the equations according to Eq. \eqref{Eq: Visibilities as sum fraction}, and solve them for the two unknown visibilities. This method is effective, as demonstrated by \citet{Caspi_2015}, where they successfully solved for the hot and superhot components and tracked the evolution of these sources over time.

    One limitation of the \citet{Caspi_2015} approach is that the results are influenced by the choice of energy ranges for constructing the equation system. The user must select the appropriate energy bins at each time step. Hereby, it is crucial that each source significantly contributes to at least one of the chosen energy ranges; otherwise, the results may appear very noisy. This requirement demands considerable attention, testing, and control from the user when selecting the energy ranges, making it a somewhat subjective process. Additionally, the \citet{Caspi_2015} method disregards part of the observed data that contains valuable information, as not all energy ranges are utilized.

    In spectral component imaging, we propose using all the available energy channels measured by the HXR instrument. This approach increases the counting statistics, uses all available information that was recorded, and gets a more objective result. Additionally, this makes spectral component imaging also more applicable to automation (see the discussions in Section \ref{Discussion} below). When dealing with three unknown, spectral sources (A, B, C), the equation system can be written as
    \begin{equation}
        \left[\begin{array}{c} \nu_1(u,v, E_1) \\ .  \\ . \\ . \\ \nu_n(u,v, E_N) \end{array}\right] = 
        \left[ \begin{array}{ccc}
        f_A(E_1) & f_B(E_1) & f_C(E_1) \\
        ... & ... & ... \\
        ... & ... & ... \\
        ... & ... & ... \\
        f_A(E_N) & f_B(E_N) & f_C(E_N) \\
        \end{array}\right] \cdot \left[\begin{array}{c} \nu_A(u,v) \\ \nu_B(u,v) \\ \nu_C(u,v) \end{array}\right].
        \label{Eq: Linear Least Square Problem}
    \end{equation}
    In the more general case of $N$ energy channels and $M$ sources (with $N\geq M$) the system becomes
    \begin{equation}
        \vec{V} = \textbf{F} \cdot \vec{\nu},
        \label{Eq: Linear Least Square Problem2}
    \end{equation}
    where the vector $\vec{V}$ contains the $N$-energy dependent visibilities $\nu_i$, the $N\times M$ matrix $\textbf{F}$ the fractional contributions $f$ and the vector $\vec{\nu}$ the $M$ visibilities of the spectral component sources $\nu_X$. Note that, in practice, it is necessary to solve a linear system (Eq. \eqref{Eq: Linear Least Square Problem2}) for each $(u,v)$ angular frequency measured by the instrument.

    To address low statistics or noisy energy channels, we weight the observed visibilities according to their uncertainties. This requires error propagation for both the real and imaginary parts of the observed visibilities. For STIX, we modified the existing ground software (GSW) v0.6.1 to incorporate the uncertainties on the real and imaginary parts of the visibilities. We assume that the recorded counts are affected by Poisson noise and compression errors, and we apply error propagation using the equations provided by \citet{Massa_2023}. 
    
    Combining all points from above and considering a least-squares approach, we minimize with respect to the components of $\vec{\nu}^{re/im}$ the residuals $\vec{r}^{re/im}$ defined as
    \begin{equation}
    \begin{split}
        \vec{r^{re}} = \sum_i \left(\frac{V_i^{re}-F_{ij}\nu_j^{re}}{\sigma_i^{re}}\right)^2\;=\;(\vec{V}^{re}-\textbf{F}\vec{\nu}^{re})^T\textbf{$\Lambda_{re}$}^{-1}(\vec{V}^{re}-\textbf{F}\vec{\nu}^{re}), 
        \\
        \vec{r^{im}} = \sum_i \left(\frac{V_i^{im}-F_{ij}\nu_j^{im}}{\sigma_i^{im}}\right)^2\;=\;(\vec{V}^{im}-\textbf{F}\vec{\nu}^{im})^T\textbf{$\Lambda_{im}$}^{-1}(\vec{V}^{im}-\textbf{F}\vec{\nu}^{im}),
        \label{Eq: LSP with covariance}
    \end{split}
    \end{equation}
    where we differentiate between the real and imaginary part of the visibilities (re/im) and $\textbf{$\Lambda$}$ is the complex covariance matrix. 
    Assuming that $\textbf{F}^T\textbf{$\Lambda_{re/im}$}^{-1}\textbf{F}$ are invertible, the least-squares approach has a closed form solution given by
    \begin{equation}
        \vec{\nu}^{re/im} = (\textbf{F}^T\textbf{$\Lambda_{re/im}$}^{-1}\textbf{F})^{-1}\textbf{F}^T\textbf{$\Lambda_{re/im}$}^{-1}\cdot\vec{V}^{re/im} := \textbf{C}^{re/im}\cdot\vec{V}^{re/im}.
        \label{Eq: Solution LSP with covariance}
    \end{equation}
    We individually construct and solve Eq. \eqref{Eq: LSP with covariance} and \eqref{Eq: Solution LSP with covariance} for both the real and the imaginary parts of the visibilities as indicated by the labels ``re'' and ``im''. The output of the linear system are $M$ visibilities for a $(u,v)$-point, each containing the morphological information of an individual spectral component. 

    Finally, to derive the uncertainty of the visibilities of the spectral components, we independently apply error propagation 
    \begin{equation}
        \sigma_{\nu_i}^{2;\; (re/im)} = \sum_j C_{ij}^{2;\; (re/im)}\;\sigma_{V_j}^{2;\; (re/im)}
        \label{Eq: Error propagation}
    \end{equation}
    to the real and imaginary parts of the visibilities.

    \subsection{Measure of the reduced $\chi^2$} \label{Def chi2 for method}

    To quantify the results of the spectral component imaging method, we measure the reduced $\chi^2$ defined by: 
    \begin{equation}
        \chi^2 = \frac{1}{n_{\rm free}-1}\sum_{i\;(E)}\sum_{j\;(u,v)}\frac{|\;\textbf{V}_{\rm obs}\;((u,v)_j,E_i)-\textbf{V}_{\rm pred}\;((u,v)_j,E_i)\;|^2}{\sigma_{amp}^2\;((u,v)_j,E_i)} ~,
        \label{Eq: chi2 visibilites}
    \end{equation}
    where $\sigma_{amp}$ refers to the uncertainty on the visibility amplitude.
    Here, we consider the sum over all considered energy channels (max. 30 for STIX) and all visibilities (typically 24, max. 30 for STIX). The uncertainty $\sigma_{amp,i}^2$ is given by $\sigma_{amp,i}^2 = \sigma_{stat}^2+\sigma_{syst}^2$, taking into account the statistical error $\sigma_{stat}$, which assumes Poisson noise and compression errors in the recorded counts, and a systematic error $\sigma_{syst}$, set at 3\% of the counts for STIX. To estimate the predicted visibilities for each energy channel, we use two different approaches. 

    The first approach involves a direct test of Eq. \eqref{Eq: Linear Least Square Problem2}. Using the solution $\vec{\nu}$ provided by Eq. \eqref{Eq: Solution LSP with covariance}, we can calculate $\textbf{V}_{pred}$ through Eq. \eqref{Eq: Linear Least Square Problem2}. This predicted value is then used as input for Eq.  \eqref{Eq: chi2 visibilites}. We refer to this as the $\chi^2$ of the linear system. 

    In the second approach, we use the reconstructed maps of the spectral components to estimate the visibilities of the spectral components. These estimates are then inserted into Eq. \eqref{Eq: Linear Least Square Problem2} to determine $\textbf{V}_{pred}$ for Eq. \eqref{Eq: chi2 visibilites}. This measure incorporates the reconstruction of the maps, which is why we will refer to it as the $\chi^2$ of the maps.

    \subsection{Albedo component}
    In this section, we discuss the impact of albedo emission on spectral component imaging. The albedo source is generated by photons from the primary source that are emitted towards the Sun and then are Compton back-scattered by the lower layers of the solar atmosphere toward the observer \citep[e.g., ][]{Kontar_2006}. In HXR spectral analysis, the albedo component is typically fitted \citep[e.g., ][]{Kasparova_2007} using the approach described by \citet{Kontar_2006}. In standard imaging techniques for indirect imaging systems, the reconstructed images contain both components, the primary source and the albedo patch. Separating the two components is challenging and appears to exceed the dynamic range of the imaging capabilities of indirect imaging, HXR instruments \citep[e.g.,][]{Kontar_2010, Battaglia_2011}.

   \begin{figure*}
   \centering
        \includegraphics[width=18cm]{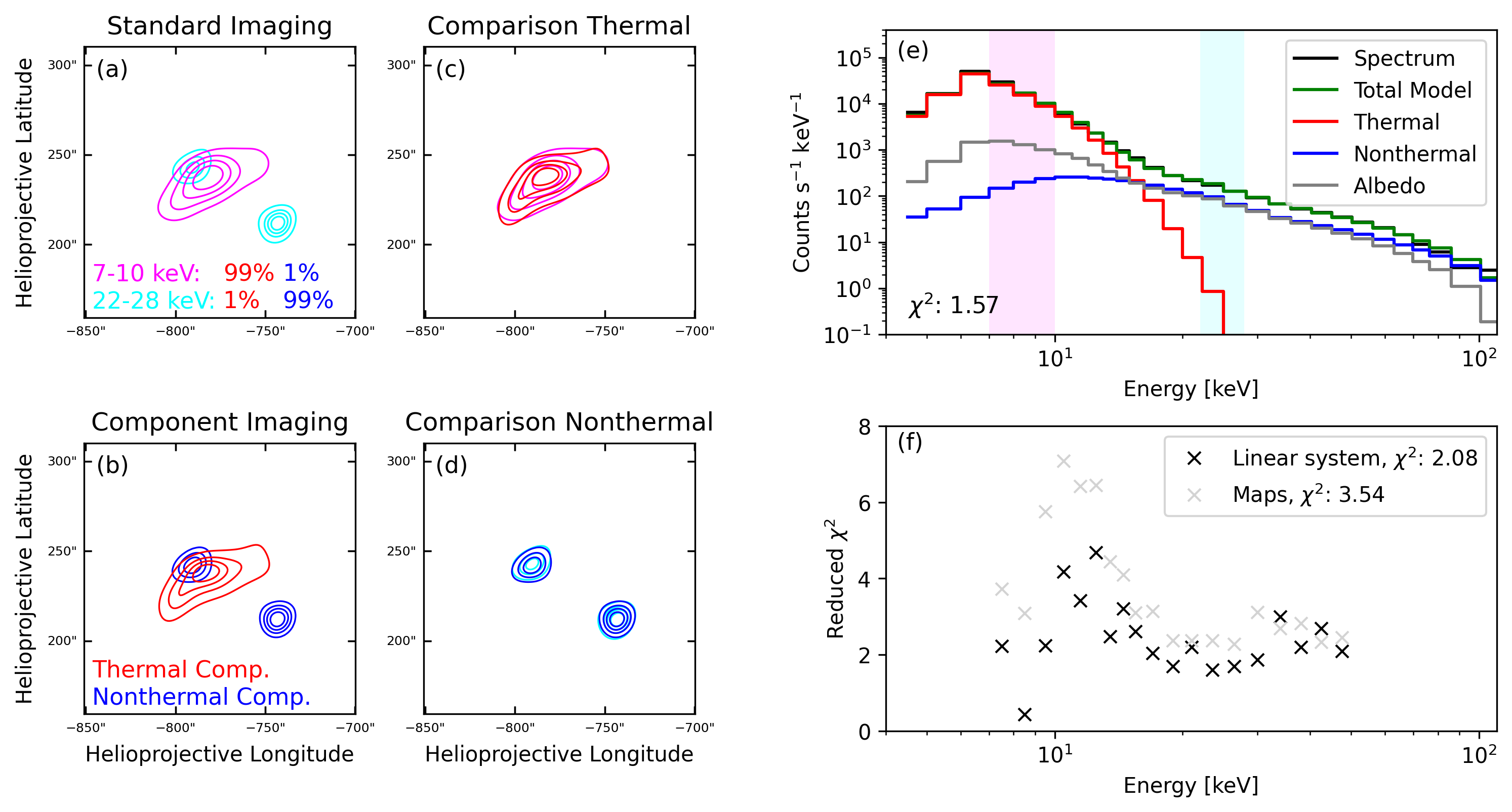}
   \caption{Panel (a) shows the reconstructed maps of the energy ranges 7-10 keV (magenta) and 22-28 keV (light blue). The percentages give the contribution of the spectral components, thermal in red and nonthermal in blue, respectively. Panel (b) shows the reconstructed maps of the thermal (red) and nonthermal (blue) spectral components as derived from spectral component imaging. Panels (c) and (d) show the comparison of pre-determined energy ranges with the spectral component imaging results for the thermal emission, panel (c), and the nonthermal emission, panel (d). The contour levels shown in all panels are 20, 40, 60, and 80\% of the peak values of the maps. Solar Orbiter was at a distance of 0.69 AU to the Sun. Panel (e) shows the spectrum of the time step used for imaging together with the fitted models. Panel (f) shows the $\chi^2$ results for the spectral component imaging approach as a function of energy. The $\chi^2$'s in the legend are the values averaged over all energies.}
    \label{Fig: Simple case}
    \end{figure*}
    
    The same difficulty of separating out the albedo component remains for spectral component imaging. In the examples presented below, we include the albedo component within the spectral analysis. To determine the fractional contributions of the thermal and nonthermal models, we subtract the albedo from the total flare emission. This method distributes the albedo emission across each spectral component according to their fractional contributions. Hence, the albedo of the footpoints is included in the resulting footpoint image. The albedo component cannot be included as an individual source in the spectral component imaging approach (in Eq. \eqref{Eq: Linear Least Square Problem}) as thermal and nonthermal sources produce different albedo patches and the combined albedo image is therefore strongly energy dependent \citep[e.g., ][]{Kontar_2006}. This energy dependency conflicts with the assumption made in our method, which is that the spectral sources are energy independent. 
    
    Considering that footpoints typically occur only around one Mm above the solar surface, the albedo patch is rather similar in shape and position to the primary source \citep[e.g., ][]{Kontar_2010, Battaglia_2011}, at least within the STIX angular resolution (typically 15 arcsec). Therefore, the albedo component is unlikely to significantly increase the source size or alter the position of the footpoints in the reconstructed images. For thermal sources, which generally originate at higher heights above the solar surface than footpoints, the albedo patch can differ from the primary source. While for near disk center sources the difference manifests in source size, near limb sources have albedo patches which are radially displaced from the primary source. Since albedo is weaker at thermal energies, and even more so for flares near the limb, the impact is somewhat suppressed. Nevertheless, one should be aware that the albedo contribution influences the derived images shown in the following. In summary, the albedo component complicates spectral component imaging, much like it does in HXR imaging in general. With the current-day instrumentation, we are unfortunately unable to properly image the albedo component, and spectral component imaging does not resolve this issue.

\subsection{Spectral component imaging using STIX data}

    For spectral component imaging with STIX, level 1 science files need to be downloaded from the STIX data center\footnote{\stixdatacenter} \citep{Xiao_2023} or the Solar Orbiter Archive (SOAR)\footnote{\solarorbiterarchive}. For the spectral analysis, OSPEX (Object Spectral Executive), the standard tool for HXR instruments in IDL\footnote{\ospex}, or SUNKIT-SPEX\footnote{\sunkitspexstix}, a spectroscopy package currently under development in Python (e.g. \citet{Bajnokov_2024}), can be used. For analyzing large flares with an inserted attenuator, we recommend using SUNKIT-SPEX along with the method of jointly fitting the STIX background detector \citep[BKG; ][]{Stiefel_2025} with the imaging detectors. The BKG detector measures flare relevant emission in the low-energy channels of STIX when the attenuator is inserted. This helps to better constrain the thermal emission in large flares. For a detailed discussion on the advantages of joint fitting, we refer to \citet{Stiefel_2025b}.

    For spectral component imaging with STIX, we developed the function ``stx\_spectral\_component\_imaging.pro'' in IDL \footnote{Adaptation of this function for RHESSI visibility data is straightforward.}.  This function will be included in a future version of the GSW of STIX. It takes as input the fractional contribution matrix \textbf{F} (Eq. \eqref{Eq: Linear Least Square Problem2}) obtained from spectral analysis, along with the STIX science file of the flare. The outputs of this function are the visibility sets for each spectral component. Using standard reconstruction methods for STIX, such as CLEAN \citep{Hogbom_1974,Dennis_2009}, MEM\_GE \citep{Massa_2020}, or forward-fit \citep{Volpara_2022}, we can reconstruct images of the individual spectral components from the corresponding visibilities. All the reconstructed images presented in this paper were obtained using MEM\_GE. There were no significant or unexpected differences between the CLEAN and MEM\_GE reconstruction methods.
    
\section{Applications of spectral component imaging on STIX data} \label{Applications}

    In this Section, we will apply spectral component imaging to four flares observed by STIX and demonstrate that the method works as intended. Spectral component imaging is particular effective for individually imaging superhot and hot components. We will demonstrate this in the example of an X7.1 GOES-class flare.

\subsection{Analysis of thermal and nonthermal emissions}

    As an initial proof of concept for spectral component imaging, we present a flare with a straightforward and well-understood structure. In this case, we can effectively separate thermal and nonthermal emissions by applying the standard method of imaging using specific energy ranges. This approach allows us to compare the results of spectral component imaging with those obtained from standard imaging, which serve as a ground truth.

    The flare shown in Fig. \ref{Fig: Simple case} is an estimated\footnote{The GOES classes are estimated for flares on the backside of the Sun using the counts in the STIX BKG detector according to \citet{Stiefel_2025}.} M5 GOES-class flare that occurred on 16 May 2023, around 17:20 UTC. In the energy range 7-10 keV, thermal emission is clearly dominant, while the 22-28 keV range is dominated by nonthermal emission, as visible in Fig. \ref{Fig: Simple case} (e). Figure \ref{Fig: Simple case} (a) shows the imaging results of these energy ranges. The flare shows a simple structure, featuring two nonthermal footpoints and thermal emission located between them, as described by the standard model for solar flares.

    We now compare the reconstructed images using the standard imaging approach with those from spectral component imaging, which uses the energy range 7-50 keV with the native binning of the energy channels of STIX. The results of spectral component imaging are presented in Fig. \ref{Fig: Simple case} (b). Additionally, the comparison between the standard imaging and spectral component imaging is shown in Fig. \ref{Fig: Simple case} (c) and (d). As visible from the contour plots, both methods yield very consistent results, demonstrating that spectral component imaging performs as expected. By normalizing the maps from both methods and comparing the pixel values - specifically (stand. imag. - spec. imag.)/(stand. imag.) - we find an average difference and a standard deviation within the 50 \% contours of -5.1 $\pm$ 6.4 \% for the nonthermal maps and -0.01 $\pm$ 25.7 \% for the thermal maps.

    Figure \ref{Fig: Simple case} (f) shows the reduced $\chi^2$ values as a function of energy for the linear system (see Eq. \eqref{Eq: Linear Least Square Problem2}) in black. This works well with a slightly enhanced $\chi^2$. Adding the reconstructed maps (see Section \ref{Def chi2 for method}) increases the difference between the predicted and observed visibilities as expected. Around 12 keV, the $\chi^2$ is clearly enhanced for this flare, possibly indicating that the source configuration is more complex than a single isothermal source. 

    In the following, we will explore two types of flares where spectral component imaging can enhance both the image quality and interpretation of these flares. 

   \begin{figure*}
   \centering
        \includegraphics[width=18cm]{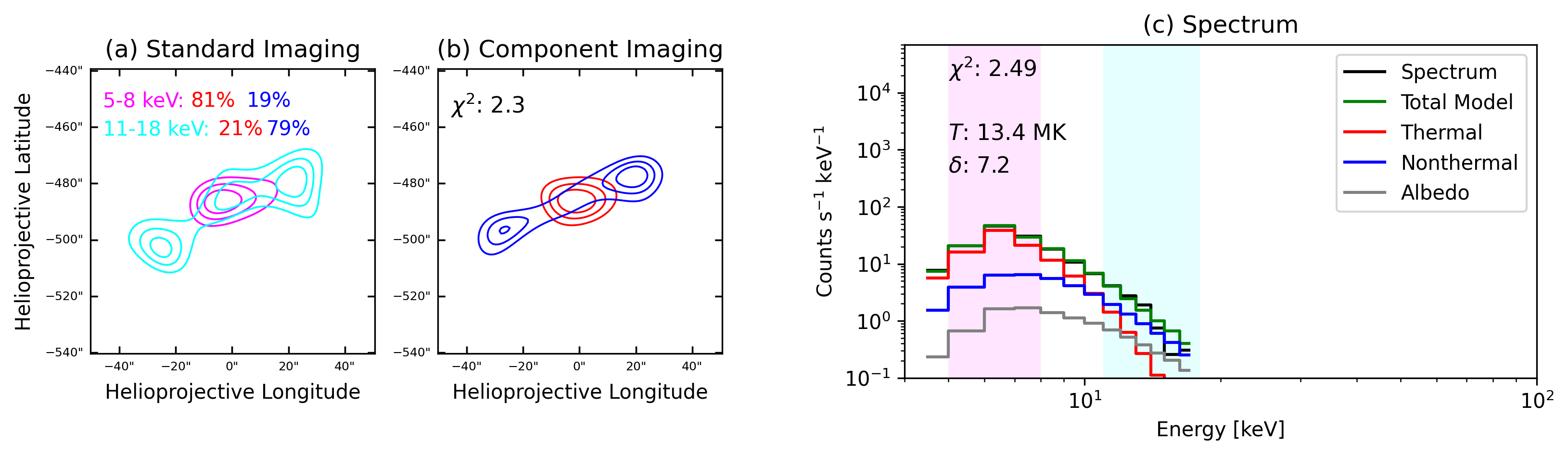}
   \caption{Results using specific energy ranges and spectral component imaging for SOL2023-03-07, with Solar Orbiter at a distance of 0.68 AU to the Sun. In panel (a), the maps using the energy ranges 5-8 keV in magenta and 11-18 keV in light blue are shown. These energy ranges are also marked in the spectrum shown in panel (c). In panel (b), the maps of spectral component imaging are shown, solving for a thermal (red) and nonthermal (blue) component, together with the map-$\chi^2$ of the spectral component imaging method. The contour levels shown in the plots are 40, 60, 80\%. In panel (c), the spectrum at the time of the image is given. We have fitted a thermal (red), nonthermal (blue), and albedo (grey) component to the spectrum. The temperature of the thermal fit, the power-law index of the nonthermal distribution and the $\chi^2$ of the fit is given in panel (c) as well.}
    \label{Fig: Multiple case}
    \end{figure*}

   \begin{figure*}
   \centering
        \includegraphics[width=18cm]{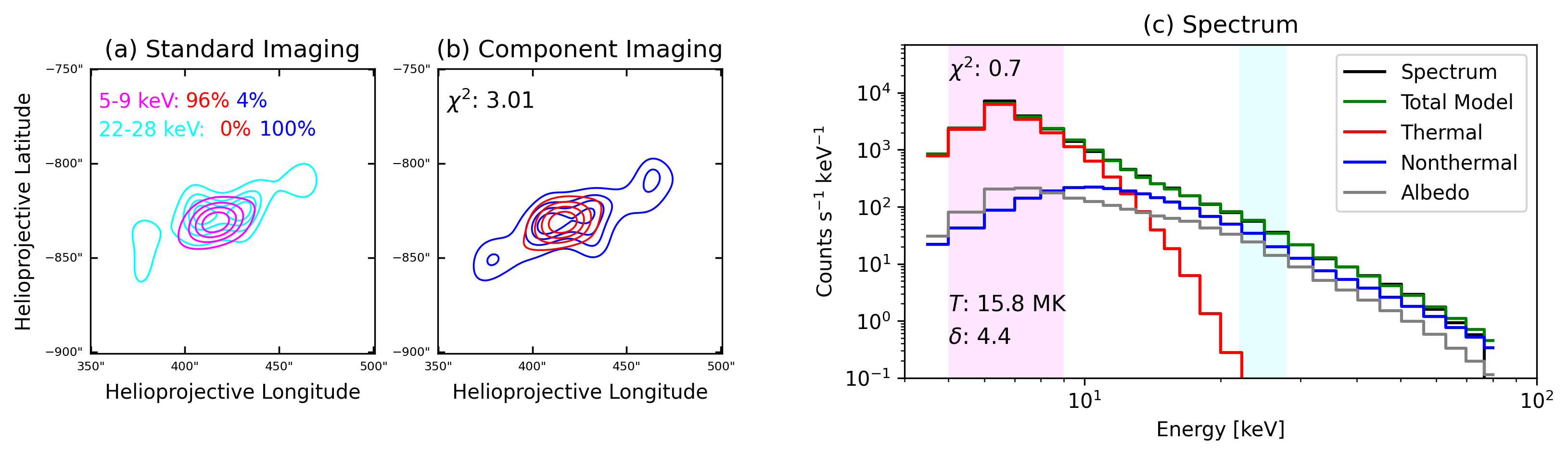}
   \caption{Results using specific energy ranges and spectral component imaging for SOL2021-09-23, with Solar Orbiter at a distance of 0.60 AU to the Sun. In panel (a), the maps using the energy ranges 5-9 keV in magenta and 22-28 keV in light blue are shown. These energy ranges are also marked in the spectrum shown in panel (c). In panel (b), the maps of spectral component imaging are shown, solving for a thermal (red) and nonthermal (blue) component, together with the map-$\chi^2$ of the spectral component imaging method. The contour levels shown in the plots are 20, 40, 60, 80 \%. In panel (c), the spectrum at the time of the image is given. We have fitted a thermal (red), nonthermal (blue), and albedo (grey) component to the spectrum. The temperature of the thermal fit, the power-law index of the nonthermal distribution and the $\chi^2$ of the fit is given in panel (c) as well.}
    \label{Fig: Multiple case 2}
    \end{figure*}

\subsubsection{Microflares}

    Imaging microflares with STIX presents significant challenges due to the limited number of counts available. Additionally, it is often not possible to image the thermal and the nonthermal emission in separate energy ranges, as the two components largely overlap in energy \citep[e.g.][]{Hannah_2008}. An example of a spectrum from an estimated B6.6 $\pm$ 0.8 flare observed by STIX on 7 March 2023 is shown in Fig. \ref{Fig: Multiple case} (c). As shown, there are no energy ranges where nonthermal emission dominates over the thermal emission. Panel (a) shows the reconstructed images using the standard imaging approach of energy ranges. Nonthermal footpoints are visible in the higher energy band, but there is additional emission in the center that overlaps with the low-energy contours. Reconstructions using the spectral component imaging approach (panel (b)) clearly show that nonthermal emission originates solely from the outer footpoints, while the thermal emission is present between the footpoints, representing the hot flare loop. Thus, spectral component imaging provides a more accurate understanding of the positions and areas of the sources (which is, e.g., important for energy budget considerations).

   \begin{figure*}
   \centering
        \includegraphics[width=14.5cm]{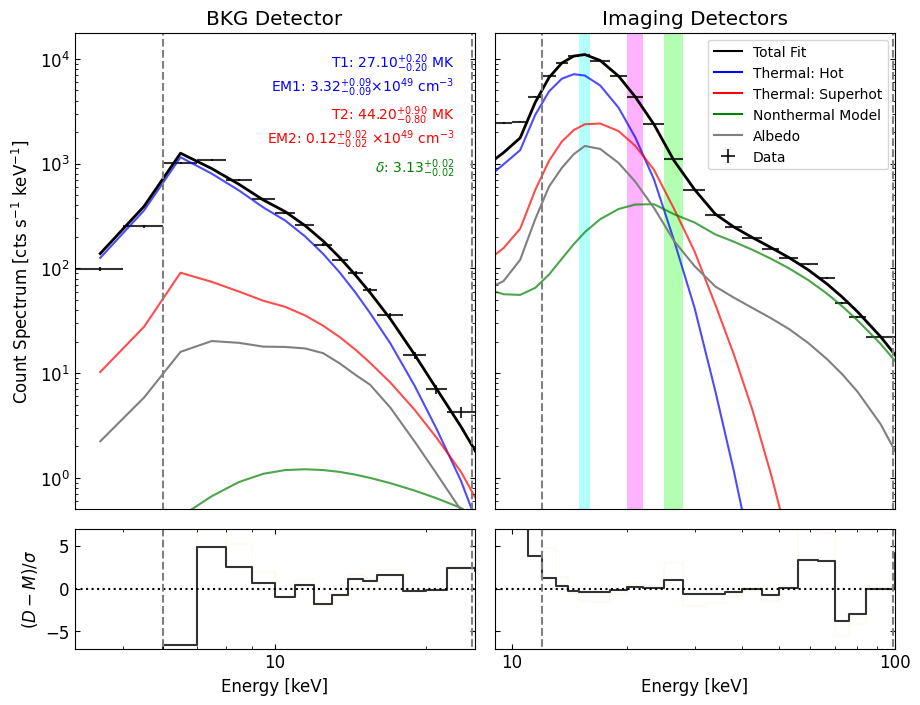}
        \includegraphics[width=18cm]{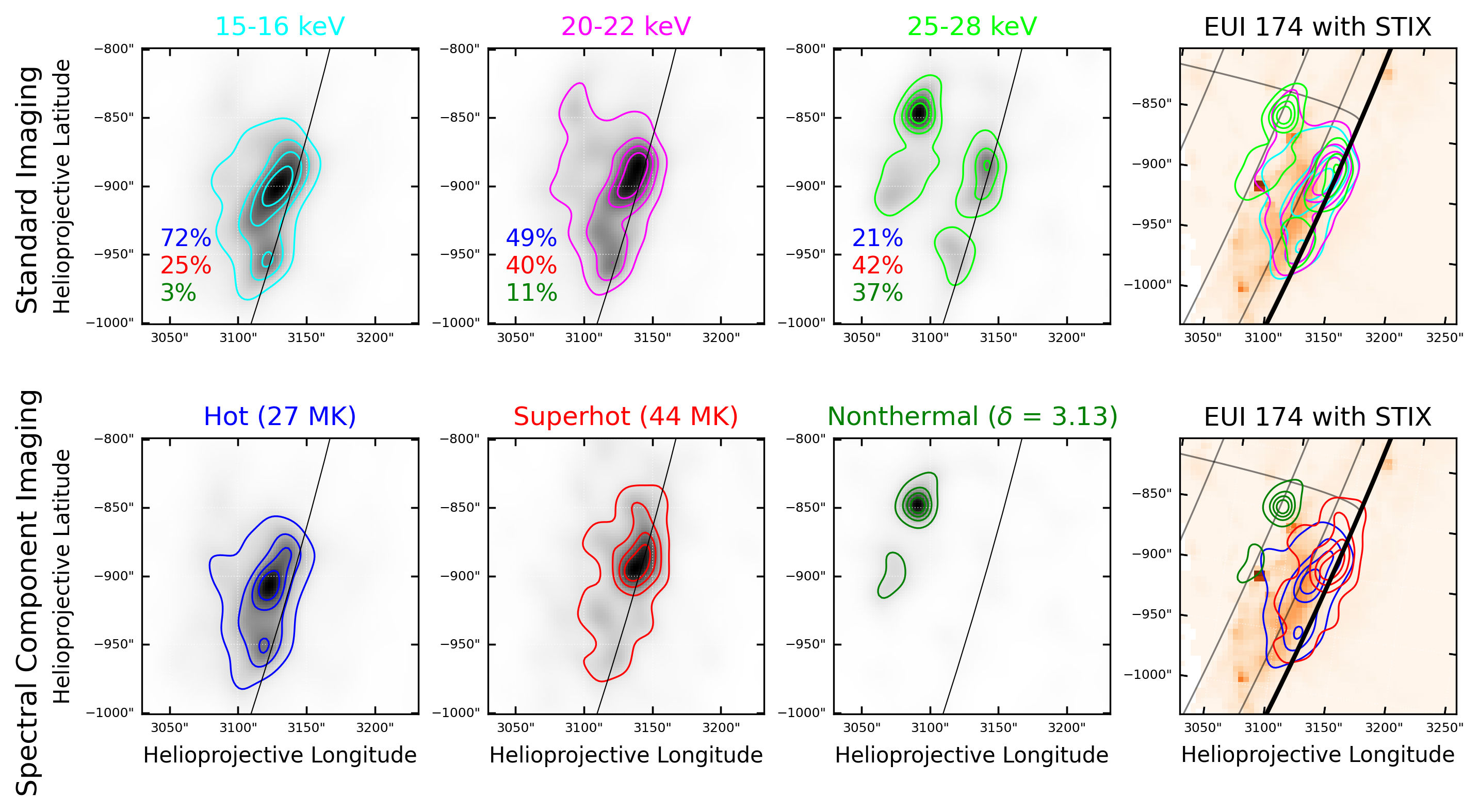}
   \caption{The top two panels show the spectra of the BKG detector (left) and the imaging detectors (right). We jointly fitted a photon model to the two spectra containing a hot thermal (blue), superhot thermal (red), nonthermal (green), and albedo (grey) component. The total fit is given in black. The ranges between the grey, dashed lines indicate the energy ranges considered for spectral fitting. 3\% systematic errors are added in quadrature to the residuals. The colored energy ranges in the spectrum are the ranges used for reconstruction and are shown in the first row of maps. For each energy range, we give the percentage of each spectral component contributing to this energy range. In the second row, we plotted the outcome of spectral component imaging, solving for the three components: hot thermal (blue), superhot thermal (red), and nonthermal (green). The contour levels of all plots shown are 20, 40, 60, 80 \% of the maximum. The plots in the right column are EUI 174~\AA{} short exposures with STIX contours. The time range shown here is 22:14:20 - 22:14:50 UTC (Earth time), a time range around the main nonthermal peak measured by STIX. Solar Orbiter is at a distance of 0.29 AU to the Sun.}
    \label{Fig: Superhot example}
    \end{figure*}

\subsubsection{Nonthermal emission at the anchor points of erupting filaments}

    \citet{Chen_2020} reported nonthermal emission in the microwave range at the anchor points of an erupting filament. This was followed by \citet{Stiefel_2023}, who, for the first time, identified nonthermal emission in the HXR range at the anchor points of an erupting filament, suggesting that nonthermal electrons can also precipitate along the magnetic field lines of the filament structure and reach the chromosphere. In the case presented by \citet{Stiefel_2023}, four distinct HXR footpoints were observed during an M1.8-class flare on 23 September 2021. The morphology of these footpoints is shown in the high-energy band in Fig. \ref{Fig: Multiple case 2}, panel (a). In a detailed multiwavelength analysis of the flare, \citet{Stiefel_2023} concluded that the two outer footpoints are likely the anchor points of the erupting filament, while the two inner sources represent the standard footpoints of the flare. To date, no similar observation of nonthermal emission at anchor points of an erupting filament in the HXR range has been reported.

    Due to the unique morphology of the flare, we used it as a test case for the spectral component imaging method. The outcomes of this method are shown in Fig. \ref{Fig: Multiple case 2}, panel (b). Spectral component imaging successfully reconstructs all four footpoints reliably. Additionally, the outer footpoints associated with the filament eruption are even slightly more distinctly defined compared to the standard imaging method shown to the left.

    As discussed in \citet{Stiefel_2023}, the reason why the four footpoints are clearly visible in the nonthermal reconstruction of this flare may be related to the similarity in intensities of the electron spectra across all footpoints (i.e., their power-law indices are similar). The absence of additional observations of HXR emission at anchor points of filament eruptions, as reported in \citet{Stiefel_2023}, can be explained by the theory that the HXR spectrum at the filament anchor points is generally softer than that at the flare footpoints. Lower energy ranges, particularly between 15-20 keV, may provide further insights into the outer footpoints. However, these energy ranges are typically not used in the standard imaging technique because they contain a mixture of nonthermal and thermal emissions, complicating the analysis. Spectral component imaging has the potential to incorporate these energy ranges, which may explain the observation that the outer footpoints are more distinctly visible when using this imaging method. Consequently, spectral component imaging could significantly aid in detecting HXR emissions at the anchor points of filament eruptions.

\subsection{Analysis of the hot and superhot thermal components}

    In this Section, we demonstrate how spectral component imaging can be used to separately image the hot and superhot thermal components. As a specific example, we focus on the X7.1 flare that occurred on the 1st of October 2024, at approximately 22:00 UTC. This flare was observed on the solar disc from Earth's perspective and near the west limb from Solar Orbiter's perspective. Due to the flare being outside of the main field of view for STIX \citep[see][]{Massa_2023}, only the top pixels of all the STIX detectors should be used for imaging and spectral analysis. 

    We show in Fig. \ref{Fig: Superhot example} how spectral component imaging enhances our understanding of flare structures that involve both nonthermal and multiple thermal components. The example shown here is during the main nonthermal peak of the flare with an imaging integration time of 30s. Due to the inserted attenuator, we used joint fitting for the spectral analysis \citep[see ][]{Stiefel_2025b} to achieve a more accurate fit of the hot and superhot components. The spectra from the two detector configurations - the BKG detector and the 24 imaging detectors with coarsest angular resolution - are shown in the top two plots of Fig. \ref{Fig: Superhot example}. We fitted two thermal components (hot and superhot), a nonthermal component, and an albedo component to both the spectra simultaneously. For the albedo component, we used an angle of 78 degrees, based on the flare's location from the perspective of Solar Orbiter. Beneath the spectra, we show the reconstructed maps for three different energy ranges: 15-16, 20-22, and 25-28 keV; and the results of spectral component imaging using the energy range 15-56 keV for the three components: hot, superhot, and nonthermal. For this time step, we obtained a $\chi^2$ = 0.85 for the linear system and $\chi^2$ = 1.42 for the reconstructed maps from the spectral component imaging routine. The panels in the right column of Fig. \ref{Fig: Superhot example} (bottom part) present a short exposure map \citep{Collier_2024} recorded by the Extreme Ultraviolet Imager \citep[EUI; ][]{Rochus_2020} in the 174~\AA{}~wavelength band. The alignment of STIX with EUI is done using the reprojected nonthermal image from HXI. We selected HXI for the alignment because it provides, in general, a more accurate pointing than STIX \citep[e.g., ][]{Warmuth_2020, Su_2024}.

    In Figure \ref{Fig: Superhot example}, it is evident that images using specific energy ranges often show significant contributions from multiple components, as it is particularly evident in the 25-28 keV energy band, which complicates the interpretation. Furthermore, there is no specific energy range that is exclusively dominated by the superhot component, making it impossible to image this component using the standard imaging approach. 

    It is important to note that fitting a single isothermal model, or two isothermal models for X-class flares, is a simplification commonly used in HXR analysis. This simplification does not fully represent the complex nature of solar flares \citep[e.g.,  ][]{Caspi_2014b}. Several differential emission measure analyses in the EUV and SXR range \citep[e.g., ][]{McTiernan_1999, Hannah_2013} have demonstrated the multi-thermal nature of solar flares. HXR analysis provides us with a single temperature for the hot component as well as for the superhot component. However, both components are likely multi-thermal. The question remains whether we can classify them as two separate plasma populations or if the two temperatures merely represent the averages of plasma temperatures within the same population. Spectral component imaging will enhance our understanding of these plasma structures by allowing us to localize them and compare their properties, including their temporal evolution.

\subsubsection{The superhot component: location and thermal energy estimate}

    In the bottom row of Fig. \ref{Fig: Superhot example}, we observe that the superhot component is located further away from the footpoints compared to the hot component. This finding aligns with previous reports \citep[e.g.,][]{Caspi_2015}. By determining the centroid positions of the hot and superhot reconstructions obtained from spectral component imaging, we can estimate the absolute distance between the centroid positions of these two sources, which is $\sim$ 4.8 Mm. In Table \ref{table: hot vs superhot} we compare the centroid position R$_{c}$ to the radius of the Sun $R_{limb}$. The difference in the radial direction between the two centroids is $\sim$1.9 Mm, with the superhot component being closer to the limb. The higher location of the superhot source supports the idea that the superhot component could be produced by direct heating related to the coronal energy release site, which is expected to be above the hot loops.

{\renewcommand{\arraystretch}{1.2}
\begin{table}
\caption{Comparison between the hot and superhot component.}
\centering
\begin{tabular}{|l|l|l|}
\hline
  &  &  \\
  & Hot  & Superhot \\
  &  &  \\
\hline
 T [MK] & 27.1 $\pm$ 0.2 & 44.2 $\pm$ 0.9 \\
 EM [10$^{49}$ cm$^{-3}$] & 3.32 $\pm$ 0.09 & 0.12 $\pm$ 0.02 \\
 $\sim$R$_{c}$-R$_{limb}$ [Mm] & -3.4  &  -1.5 \\
 V [10$^{26}$ cm$^{3}$] & 7.77  & 3.95 \\
\hline
 $\rho$ [10$^{11}$ cm$^{-3}$] & 2.1  & 0.6  \\
 E$_{th}$ [10$^{30}$ erg] & 1.8  & 0.4  \\
 \hline
\end{tabular}
\tablefoot{This table summarizes the comparison of the physical properties between the hot and superhot thermal components, given in the left and right columns, respectively. The temperature $T$ and emission measure $EM$ are the results of the spectral fitting shown in the top plots of Fig. \ref{Fig: Superhot example}. The distance of the centroid to the solar limb, R$_{c}$-R$_{limb}$, and the volume $V$ are determined using the reconstructed images from spectral component imaging. The density $\rho$ and the thermal energy E$_{th}$ are determined using Eq. \eqref{Eq: density} and \eqref{Eq: thermal energy}. The time range considered here is the same as for Fig. \ref{Fig: Superhot example}, 22:14:20-22:14:50 UTC (Earth time).}
\label{table: hot vs superhot}
\end{table}
}

To date, determining the relative thermal energy content of a flare for the hot and the superhot components is missing. \citet{Caspi_2014} provides some estimates for the thermal energy content of the superhot component, but does not compare it to the hot component measured by the HXR instrument. Using the reconstructions shown in Fig. \ref{Fig: Superhot example}, we can independently determine the thermal energy of the hot and superhot components associated with the nonthermal peak. To estimate the volume from 2D images, we use the simplest approach in HXR imaging \citep[e.g.][]{Warmuth_2013, Ryan_2024}. We extract the area $A$ within the 50\% contours of the maps and use the formula $V = A^{3/2}$ to estimate the volume. Using this volume $V$ and the temperature $T$ and emission measure $EM$ from the spectral fitting, we can calculate the thermal energy $E_{th}$ using the equation
\begin{equation}
    E_{th} \; = \; 3k_BT\sqrt{EM\;V},
    \label{Eq: thermal energy}
\end{equation}

and the plasma density $\rho$, assuming a filling factor of 1, using the equation

\begin{equation}
    \rho \; = \; \sqrt{\frac{EM}{V}}.
    \label{Eq: density}
\end{equation}
The results are presented in Table \ref{table: hot vs superhot}. It is important to note that the estimation of densities and thermal energy content carry a significant degree of uncertainty due to the volume estimate derived from a 2D image. More advanced techniques, such as 3D imaging, could provide better estimates, although these methods also present their own challenges \citep[][Palumbo et al., in prep.]{Ryan_2024}. The derived values given in Table \ref{table: hot vs superhot} are within the range of previously observed flares \citep[e.g., ][]{Caspi_2014, Warmuth_2016, Stiefel_2025b}. 
The density estimates of the superhot component of $0.6\times10^{11}$ cm$^{-3}$ is an average value compared with the statistical study of \citet[][]{Caspi_2014}. The hot loop is denser, as expect in a scenario where heated chromospheric plasma creates the hot loops.

The energy content of the superhot component is approximately 22\% of that in the hot component during the nonthermal peak. This finding corroborates the order of magnitude estimates made in \citet{Stiefel_2025b} and emphasizes the significance of the superhot component in the total thermal energy of the flare. The ratio may vary during the decay phase and needs to be studied in a detailed temporal analysis of the superhot component's evolution.

\subsubsection{Final comments on the analysis of the superhot component}

This paper aimed to demonstrate how spectral component imaging with STIX data can effectively image the hot, the superhot, and the nonthermal component individually. We show that spectral component imaging, together with joint spectral fitting \citep{Stiefel_2025b}, serves as a powerful diagnostic tool for conducting detailed temporal analyses of the thermal emissions using STIX data. In our future work, we plan to investigate the physical properties of large X-class flares, with a particular focus on the superhot component, using spectral component imaging.  

\section{Discussion and conclusion} \label{Discussion}

In this paper, we introduced a new method called ``spectral component imaging''. This method determines the visibilities of individual spectral components (e.g., thermal and nonthermal components), rather than relying on the visibilities recorded within specific energy ranges. The key assumption we made for spectral component imaging is that the spatial morphologies of the sources corresponding to the spectral components are energy independent.

We demonstrated that our method performs as anticipated using STIX data and provided several examples showing how spectral component imaging can enhance imaging with indirect HXR instruments. Spectral component imaging delivers unmatched results for the analysis of large flares, in particular for separately imaging the hot and superhot components. 

The advantage of spectral component imaging lies in its inclusion of all energy bins that contain signals of multiple spectral sources in the analysis. In typical C/M-class flares, the standard imaging method is often well justified, as all relevant information is found in the distinct lower and higher energy bands (as demonstrated in the flare in Fig. \ref{Fig: Simple case}). However, this paper shows that there are several types of flares where spectral component imaging improves our understanding of the flare's morphology. 

Spectral component imaging, like any imaging method using indirect imaging systems, is limited by statistics. Since this method relies on the native energy bins of STIX, it requires a sufficiently high overall number of counts in all energy bins used in the process. Despite this limitation, we demonstrated that spectral component imaging can be applied across a wide range of flares, from a small B7 class flare to a large X7 class flare. While spectral component imaging is limited by the dynamic range of STIX, the method is effective in differentiating between two sources when they originate from different spectral components, i.e., of different physical origins. 

The main drawback of the presented methodology is that it requires three distinct steps: spectral fitting, solution of linear systems for retrieving the visibilities of the individual spectral components, and image reconstruction. 
However, single-step approaches combining spectral fitting and image reconstruction are currently under development (Guidetti et al., in preparation) and will be tested and compared to the methodology presented in this paper.

To clarify the distinction between ``spectral component imaging'' and ``imaging spectroscopy'', we will briefly outline the differences in the following.

\subsection{Spectral component imaging and imaging spectroscopy}\label{IS vs SCI}

Imaging spectroscopy has been used in the past to derive spectra of individual sources seen in images using standard imaging routines \citep[e.g.][]{Krucker_2002}. The flux as a function of energy is extracted from (often manually) selected regions of interest (e.g., footpoints or flare loops). The spectrum from these individual regions can then be used for spectral analysis. This approach allows for the comparison of, e.g., the spectra of two footpoints within the same flare on a separate basis \citep[e.g., ][]{2003ApJ...595L.107E}, or to compare nonthermal sources from the chromosphere with sources in the corona \citep[e.g., ][]{2014ApJ...780..107K}.

Thus, spectral component imaging and imaging spectroscopy are fundamentally different methods. Depending on the scientific questions and the information desired, one method or the other should be applied. To compare the spectra or the time evolution between two footpoint sources, imaging spectroscopy is the approach to use. The relative fluxes of two footpoints in spectral component imaging represent an energy-averaged flux ratio with the relative spectral information lost. An example where spectral component imaging is the preferred method are flares with superhot sources. Only spectral component imaging is able to spatially separate these components, whereas imaging spectroscopy relies on the standard imaging technique in which the locations of the sources are often not well recognized.

Creating an image database of thermal and nonthermal flare sources is a further application of spectral component imaging, which is briefly discussed in the following section. 

\subsection{Automatic imaging - Level 3 data product for STIX} \label{Level 2 product STIX}

For the current database of over 85'000 flares measured by STIX (state fall 2025), automatic analyses are essential to create an image database for statistical studies or machine learning applications \citep[e.g., ][]{Selcuk_2025}. Such an image database for STIX flares would be a valuable resource for users who do not want to go into the details of how STIX imaging works. However, creating an image database using an indirect imaging system comes with various challenges. Several factors need to be carefully selected, including the time range, energy range, and selection of detectors, to be able to handle small B-class flares all the way up to the largest X-class flares in an appropriate way. 

Spectral component imaging has the potential to support this effort by removing the need to select energy ranges and directly providing images of physically meaningful sources. This is achievable because spectral component imaging can be run automatically once the number of spectral components is determined. For creating an image database, thermal and nonthermal images would be reconstructed for each flare, resulting in a data product that is relatively straightforward to understand. This is in contrast to the standard imaging approach that reconstructs images within fixed energy ranges, where interpreting the images is generally far more complex. 

Our approach discussed above has already been tested to run fully automatically once the spectral analysis is done manually. While it is technically possible to run the spectral analysis automatically (e.g., in OSPEX) and a first iteration has already been implemented to the STIX website, further refinements—such as automatic, iterative fitting— and tests are necessary to enhance the reliability and quality of the spectral fitting results. Additionally, the decision regarding which spectral components and how many to include remains open. Our suggestion is to use a thermal and nonthermal model around the nonthermal peak; however, this suggestion needs further discussion and testing.

\subsection{Conclusion}

In this paper, we introduced a new method for indirect X-ray imaging called spectral component imaging and demonstrated its effectiveness using STIX data. We highlighted specific scenarios where spectral component imaging can outperform the standard imaging method, particularly in the analysis of microflares or flares associated with filament eruptions. 

Spectral component imaging is especially effective for imaging the superhot and hot thermal components found in large X-class flares, as demonstrated on the X7.1 class flare that occurred on the 1st of October 2024. To spatially separate these two sources and analyze their physical properties, including thermal energy, spectral component imaging is essential. The initial science output of spectral component imaging corroborates that the hot and superhot components are indeed spatially separated and that the thermal energy content of the superhot source is about a fifth of the thermal energy of the hot flare loop. To do a statistical study of the properties of superhot sources, including their time evolution, is the planned follow-up project.

\begin{acknowledgements}
      \em{This work} was conducted using tools provided by sunpy \citep[][]{sunpy_community2020}, astropy \citep[][]{astropy:2022}, matplotlib \citep[][]{Hunter_2007}, numpy \citep[][]{harris2020array} and the SUNKIT-SPEX package. Solar Orbiter is a space mission of international collaboration between ESA and NASA, operated by ESA. The STIX instrument is an international collaboration between Switzerland, Poland, France, Czech Republic, Germany, Austria, Ireland, and Italy. PM is partially supported by the Swiss National Science Foundation in the framework of the project Robust Density Models for High Energy Physics and Solar Physics (rodem.ch), CRSII5\_193716. MS thanks the Institute for Data Science at FHNW for their continued support during her doctoral studies. The authors thank the anonymous referee for the helpful comments which improved the paper.
\end{acknowledgements}

\bibliographystyle{aa} 
\bibliography{aa57373-25} 

\end{document}